\documentclass[aip, prl, preprint,  superscriptaddress]{revtex4} 

\newcommand{\kB}{k_\mathrm{B}}

\usepackage{graphicx}
\usepackage{hyperref}
\usepackage{epstopdf}			
\usepackage{color}
\usepackage{amsmath} 
\usepackage{mathtools} 
\usepackage{ctable}
\usepackage{tabularx}

\date{\today}

\begin{document}

\title[latticefluids]
  {{Structuring of fluid adlayers upon ongoing unimolecular adsorption}}  
\date{\today}

\author{C. Schaefer}
\email{charley.schaefer@durham.ac.uk}
\affiliation{Department of Physics, Durham University, South Road, DH1 3LE, UK}

\begin{abstract}
  Fluids with spatial density variations of single or mixed molecules play a key role in biophysics, soft matter and materials science.
  The fluid structures usually form via spinodal decomposition or nucleation following an instantaneous destabilisation of the initially disordered fluid.
  However, in practice an instantaneous quench is often not viable, and the rate of destabilisation may be gradual rather than instantaneous. 
  In this work we show that the commonly used phenomenological descriptions of fluid structuring are inadequate under these conditions. 
  We come to that conclusion in the context of surface catalysis, where we employ kinetic Monte Carlo simulations to describe the unimolecular adsorption of gaseous molecules onto a metal surface.
  The adsorbates diffuse at the surface and, as a consequence of lateral interactions and due to an ongoing increase of the surface coverage, phase separate into coexisting low- and high-density regions.
  The typical size of these regions turns out to depend  much more strongly on the rate of adsorption than predicted from recently reported phenomenological models.
  {We discuss how this finding contributes to the fundamental understanding of the crossover from liquid-liquid to liquid-solid demixing of solution-cast polymer blends.}
\end{abstract}


\maketitle


A common theme in the fields of soft matter and surface-catalysed chemical reactions is the crucial role of mesoscopic phase separation: 
While soft matter deals with topics such as the demixing of lipids in bilayer membranes  \cite{McCarthy15} and of polymers or nanoparticles in a melt or solution \cite{Binder83, Rabani03, Love05, VanRoekel13, Franeker15A, Chalmers17}, surface catalysis relies on the structure of coexisting low- and high-density fluid domains of reactants adsorbed at a metal surface \cite{Ziff86, Zhdanov97, KatsoulaksisBook04, VanSanten09, JansenBook12, Schaefer13, Stamatakis15, Pineda17}. 
{In this work, we theoretically explore under what conditions phenomenological models fail to describe phase separation in the context of surface catalysis, and discuss the implications of this failure on the morphology formation in solvent-cast thin-film polymer composites  \cite{Jukes05, Cao15, Franeker15B, Franeker15C, Schaefer16, Schaefer17}.}

{
In particular, we are interested in the mechanism that determines whether evaporation-induced phase separation of the polymer blend in solution takes place via liquid-liquid (L-L) or via liquid-solid (L-S) demixing \cite{Cao15, Franeker15B, Franeker15C}.
A change from L-L to L-S demixing results in a tremendous change of the dry-layer morphology and hence the macroscopic properties of the polymer composite, and may be achieved through the addition of a high-boiling-point cosolvent.
The underlying mechanism remains elusive \cite{Franeker15B}, and attempts to address it using molecular dynamics simulations are severely challenged by the required time scales and system sizes \cite{Tsige05, Krishnan15, Negi16}. 
As an alternative approach, we propose to seek this crossover under conditions where phenomenological models to describe L-L demixing break down.
We identify such conditions for the case where the concentration increases uniformly, i.e., in the absence of stratified concentration profiles \cite{Jukes05, Buxton07, Coveney14, Zoumpouli16, Schaefer16, Schaefer17}.
For these conditions, we expose the failure of phenomenological models by finding a rather strong dependence of the length scale of the phase-separated domains on the quench rate.
}

{
Rather than attempting to include all aspects of the detailed chemistry in computationally expensive molecular dynamics simulations, we capture the relevant physics of uniformly destabilised systems using a simple two-dimensional lattice model \cite{Cahn61, Huston66, Bray94, Schaefer15, Schaefer16}, evaluated using kinetic Monte Carlo (kMC) simulations \cite{Carmesin86, Puri94, Glotzer95,Puri98}.
Previous works showed that the time dependence of the composition alters the spinodal wavelength following an instantaneous temperature quench into the miscibility gap \cite{Puri94, Glotzer95,Puri98}.
On the other hand, if the quench is gradual rather than instantaneous, then local mean-field models predict the spinodal length scale to decrease with the one-sixth power of the evaporation rate \cite{Huston66, Nauman88, Schaefer15, Schaefer16, Kessler16}.
Although it is known that  nucleation may play a role in experimental systems \cite{Buil00, Rullman04, Lutsko12}, the validity of the mean-field approximation was not tested systematically.
}

{
A particularly suitable study case for this analysis, is the situation in surface catalysis of unimolecular adsorption of gaseous molecules onto an initially clean metal surface \cite{Zhdanov97, Ikemiya00, RodriguezBook13, ChenQ15}.
Due to ongoing adsorption the surface coverage of adsorbates increases, which leads to a composition quench that induces phase separation of the laterally interacting adsorbates.
In the following, we set up the model system and kMC scheme, and investigate the impact of the adsorption rate upon structure formation.
Finally, we discuss the consequences of the discrepancies between our findings and those expected from phenomenological models, and discuss the implications to solvent-borne thin-film polymer composites. 
}


To set up the model, we consider a metal with initially vacant adsorption sites that are ordered in a periodic $N^2$ square lattice.
In our simulations, we typically choose $N=256$ and average over multiple (typically five or more) lattices to enhance the statistics.
To keep our description as simple as possible we only allow five types of events to take place:
Gaseous molecules may adsorb at a site (but not desorb), or an adsorbed molecule may hop to one of the four nearest-neighbour sites.
Within the `Random Selection Method' in each kMC step one out of a list of a $5N^2$ events is potentially possible \cite{JansenBook12}, and time increases with an increment
\begin{equation}
  \Delta t = -\frac{\ln u}{5N^2r_\mathrm{max}},
\end{equation}
where $u$ is a uniform deviate on the unit interval and $r_\mathrm{max}$ is the upper value of an event rate.

During a kMC step, an event is randomly selected from this list and may either be immediately rejected, e.g., because it represents adsorption or hopping to a site that is already occupied, or accepted according to an acceptance probability
\begin{equation}
  P = \frac{r}{r_\mathrm{max}},
\end{equation}
where $r$ is the rate of the event.
We have specified the rates of nearest-neighbour hopping and adsorption as follows.

For the hopping rate we take \cite{Singh11}
\begin{equation}
  W_{i\rightarrow j}^{\mathrm{diff}}=\nu_\mathrm{hop} \exp (-\Delta \mathcal{H}/\kB T),
\end{equation}
where $\kB$ is Boltzmann's constant and $\nu_\mathrm{hop}(T)$ is the attempt frequency of a hop.
In principle, the latter depends on the activation barrier that the adsorbate has to overcome to escape a lattice site and hence depends on the absolute temperature, $T$ \cite{JansenBook12}. 
In order to model fluid structuring, we include lateral interactions using the Ising Hamiltonian 
\begin{equation}
  \mathcal{H} = -\frac{1}{4}J\sum_{\langle ij \rangle} \theta_i\theta_j,
\end{equation}
where the sum includes all nearest neighbours $\langle ij \rangle$.
Finally, $\theta_{i}\in \{0,\,1\}$ is the occupancy of site $i$ and $J$ is the coupling parameter.

\begin{figure}   
\centering
\includegraphics*[trim = 0.15cm 0.0cm 0.50cm 0.0cm, width=8cm]{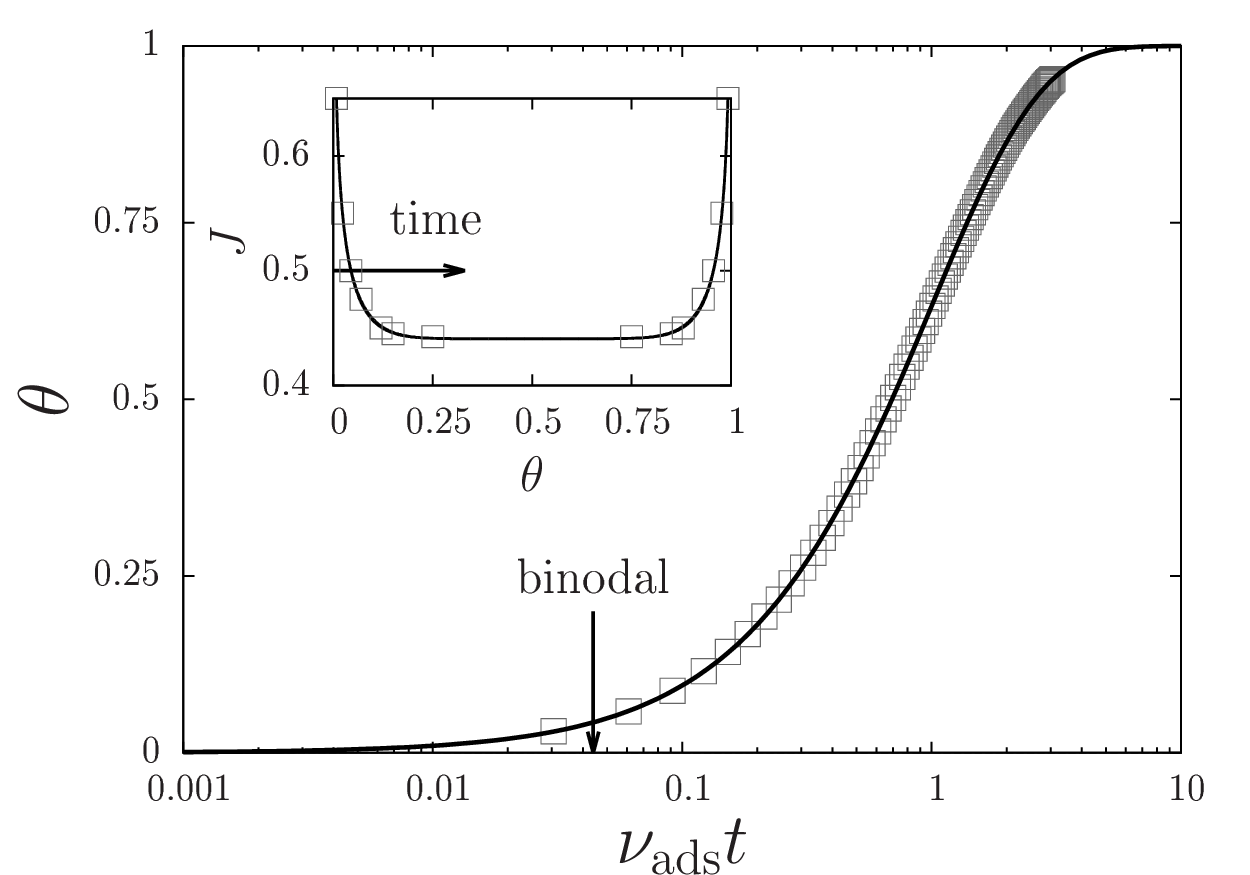}
\caption{
Surface coverage $\theta$ as a function of the dimensionless time $\nu_\mathrm{ads}t$, with $\nu_\mathrm{ads}$ the rate of adsorption and $t$ the time. 
The symbols represent all kMC results obtained in this work and the solid line is given by Eq.~(\ref{eq:adsorption_macro}). 
The inset shows how the increase in coverage implies that the unstable region of the phase diagram is entered, provided that $J$ is larger than the critical value of $0.44$. See main text.
 }
\label{fig:Coverage}
\end{figure}

If the coupling parameter is negative the adsorbates repel each other, whereas for positive values they attract each other.
For sufficiently large attraction the particles self organise into coexisting low- and high-density regions.
This happens for a coupling parameter larger than the critical value of $0.44$ \cite{HuangBook87}, as indicated by the well-known Ising phase diagram  shown in the inset Figure~\ref{fig:Coverage}.
The purity of the coexisting domains is indicated by the binodal, given by $J=(1/2)\mathrm{arsinh}([1-(2\theta-1)^8]^{-1/4})$ (solid line).
We used this analytical expression to validate our simulated results (symbols) \footnote{We have obtained the phase diagram by considering two isolated lattices that are initially fully occupied and clean, respectively. During a Monte Carlo (MC) step a random site pair was selected and swapped according to the appropriate acceptance probability.
After a sufficient number of MC steps the coverages of the two lattices converged to the coexistence values. 
}.

The dynamics by which our lattice fluid isothermally destabilises is indicated by the arrow in the inset of Figure~\ref{fig:Coverage}. 
This arrow represents the trajectory through the phase diagram as followed by our lattice fluid  for a coupling parameter of $J=0.5$: Initially, the adlayer resides in the single-phase region of the phase diagram, and enters the two-phase coexistence region when the coverage crosses the value $\theta \approx 0.04$.
To verify that our findings are robust with respect to the distance from the critical point, we have varied the value of the coupling parameter from $0.45$ to $0.76$.
The rate by which the adlayers enter the coexistence region of the phase diagram is described by the final ingredient of our model, which is the microscopic adsorption rate.

\begin{figure}   
\centering
\includegraphics*[width=2.9cm]{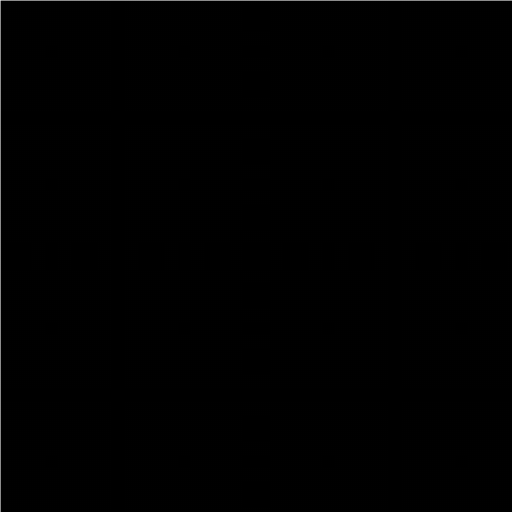}
\includegraphics*[width=2.9cm]{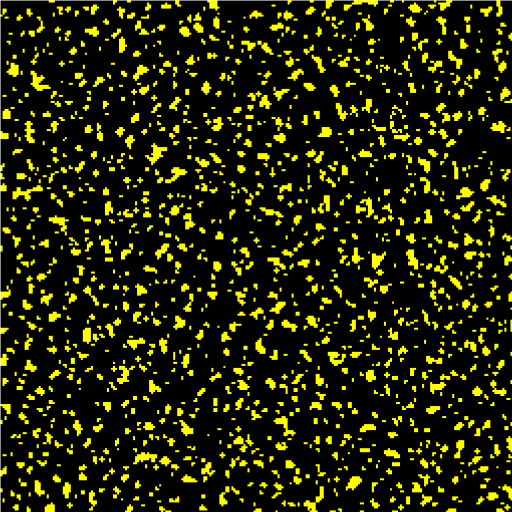}
\includegraphics*[width=2.9cm]{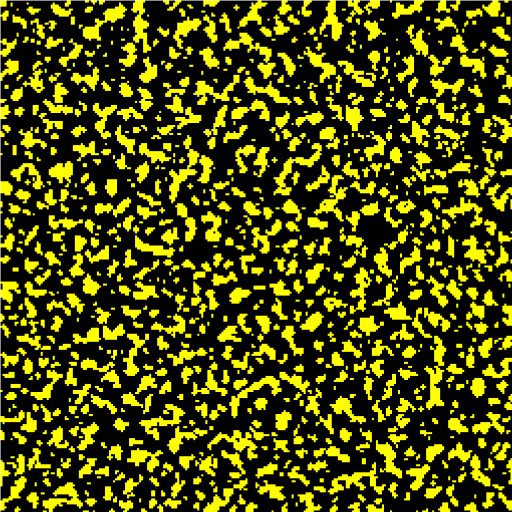}
\includegraphics*[width=2.9cm]{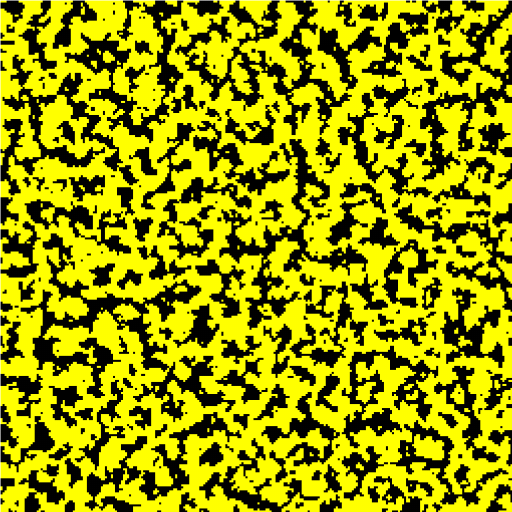}
\includegraphics*[width=2.9cm]{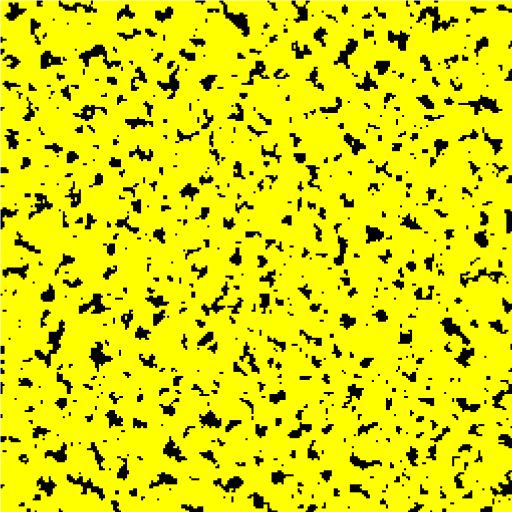}
\caption{Configurations of adsorbates at a $256\times256$ periodic lattice with a coupling parameter $J=0.5$ and a dimensionless adsorption rate of $\nu_\mathrm{ads}=0.0001 \nu_\mathrm{hop}$ for times $\nu_\mathrm{ads}t=0,0.061,\,0.122,\,0.305,\,0.610$, with $\nu_\mathrm{hop}$ the attempt frequency of a hop. The dark areas represent vacant sites, whereas the light areas are occupied by the adsorbate. }
\label{fig:Config}
\end{figure}

For reasons of simplicity, we ignore the influence of lateral interactions on the adsorption rate \cite{Kreuzer96, Lausche13} and use the unimolecular adsorption rate \cite{JansenBook12}
\begin{equation}
  W_{i\rightarrow j}^{\mathrm{ads}}=\nu_\mathrm{ads}. \label{eq:adsorption_micro}
\end{equation}
We discuss the implications of this crude approximation at the end of this Letter. 
The rate in Eq.~(\ref{eq:adsorption_micro}) results in the macroscopic adsorption rate given by
\begin{equation}
  \frac{\mathrm{d}\theta}{\mathrm{d}t}=(1-\theta)\nu_\mathrm{ads}, \label{eq:adsorption_macro}
\end{equation}
of which the solution is $\theta(t) = 1-\exp(-\nu_\mathrm{ads} t)$, provided that the surface is clean at time $t=0$. 
This relation is represented by the solid line in the main graph of Figure~\ref{fig:Coverage}.
The symbols in Figure~\ref{fig:Coverage} correspond to a simulation for a coupling parameter $J=0.5$ and adsorption rate $\nu_\mathrm{ads}/\nu_\mathrm{hop} = 10^{-4}$. 
For this simulation, we visualise the time development of the adlayer structure in Figure~\ref{fig:Config}.

Figure~\ref{fig:Config} shows an initially clean surface with exclusively vacant sites (black).
As time proceeds, an increased number of sites gets occupied by adsorbates (yellow) until the surface is completely covered. Complete coverage is achieved after a time of the order of the reciprocal adsorption rate, $\nu_\mathrm{ads}^{-1}$.
During structure evolution we observe the formation of clusters that coarsen on account of ripening and coalescence, as well as due to material addition by adsorption.
In order to characterise how the structure evolves with time, one may obtain a typical length scale from the first moment or the maximum of the radially averaged structure factor, or from the first moment, first minimum or first zero point of the radially averaged correlation function \cite{Singh11}.
Of these measures, the last one is best defined and is guaranteed to decay from unity to some negative value.
Indeed, it turns out to provide the most robust value, hence we define this as the characteristic length scale, $R_\ast(t)$ .

\begin{figure}   
\centering
\includegraphics*[width=8cm]{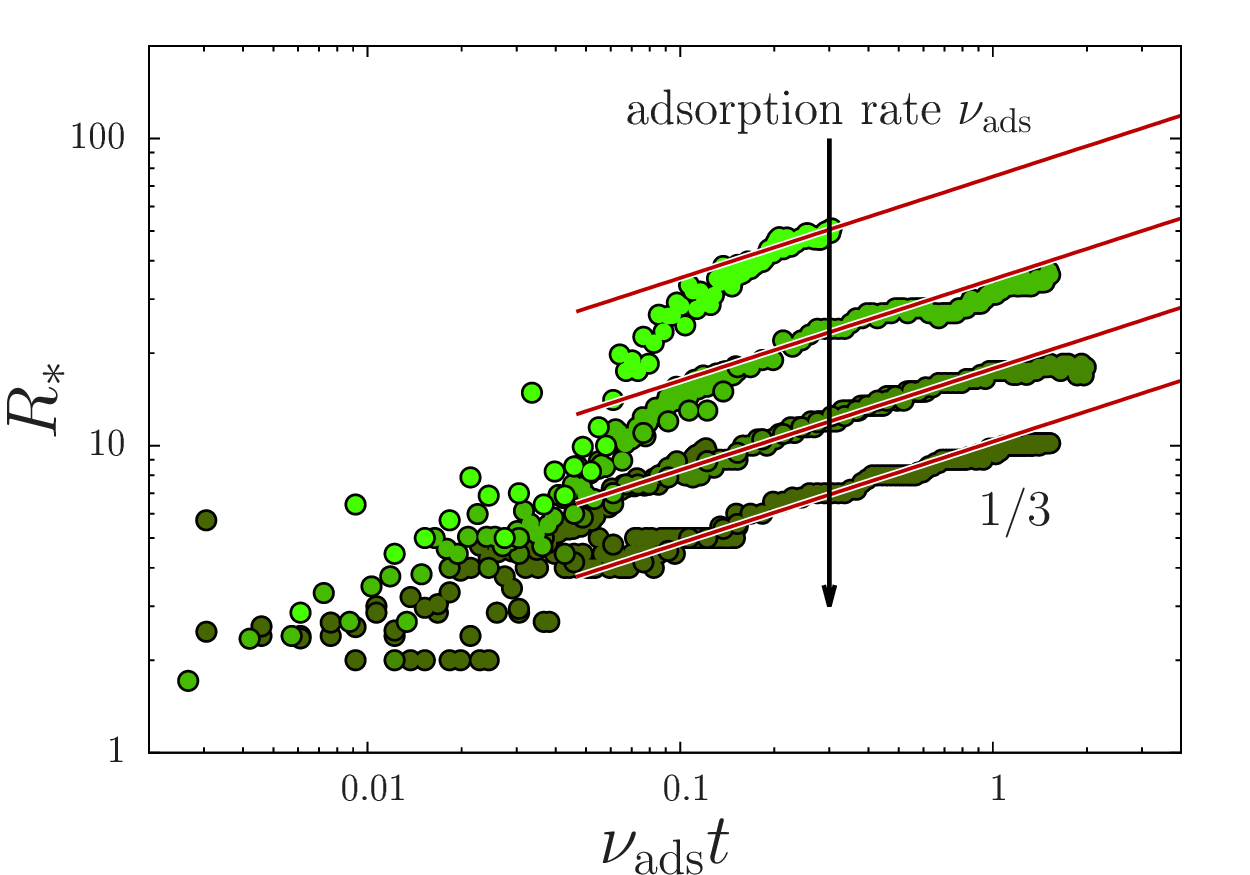}
\caption{Structural length scale, $R_\ast$, as a function of time $t$, with the time in units of the adsorption rate $\nu_\mathrm{ads}$. For long times, the structural length develops as $R_\ast=R_0(t/t_0)^{1/3}$, with $t_0$ the time at which the coverage crosses the binodal value.}
\label{fig:scaling1}
\end{figure}

It is now of interest to compare the time evolution of this length scale to what is expected from the phenomenological models \cite{Huston66, Benmouna13, Schaefer15, Schaefer16, Kessler16}.
To serve this purpose we have plotted the time evolution of $R_\ast(t)$ in Figure~\ref{fig:scaling1} for adsorption rates, $\nu_\mathrm{ads}$, ranging from $10^{-6}\nu_\mathrm{hop}$ to $10^{-3}\nu_\mathrm{hop}$.
The time is given in units of the reciprocal adsorption rate, $\nu_\mathrm{ads}^{-1}$, so that  for all curves the binodal coverage is crossed at time $\nu_\mathrm{ads}t\approx 0.04$.
For times $\nu_\mathrm{ads}t < 0.04$ we find the first discrepancies between our microscopic model and the expectations from the phenomenological models \cite{Huston66, Benmouna13, Schaefer15, Schaefer16, Kessler16}:
While the latter do not include any structuring prior to reaching the spinodal, we clearly find structuring in the single-phase region prior to entering the unstable region.

Obviously, the presence of a correlation length in the single-phase region is tiexpected \cite{Cahn61, Cook70, Langer71, Carmesin86, Lutsko12}. It does however not directly explain the behavior just after entering the unstable region of the phase diagram:
While the mean-field models predict a spinodal length that \emph{decreases} with time due to ongoing destabilisation of the mixture \cite{Benmouna13, Schaefer15, Schaefer16, Kessler16}, we here find a structural length scale that \emph{increases} with time prior to reaching the usual $R_\ast \propto t^{1/3}$ scaling at late times (solid lines) \cite{YaoJH93}.

\begin{figure}   
\centering
\includegraphics*[width=8cm]{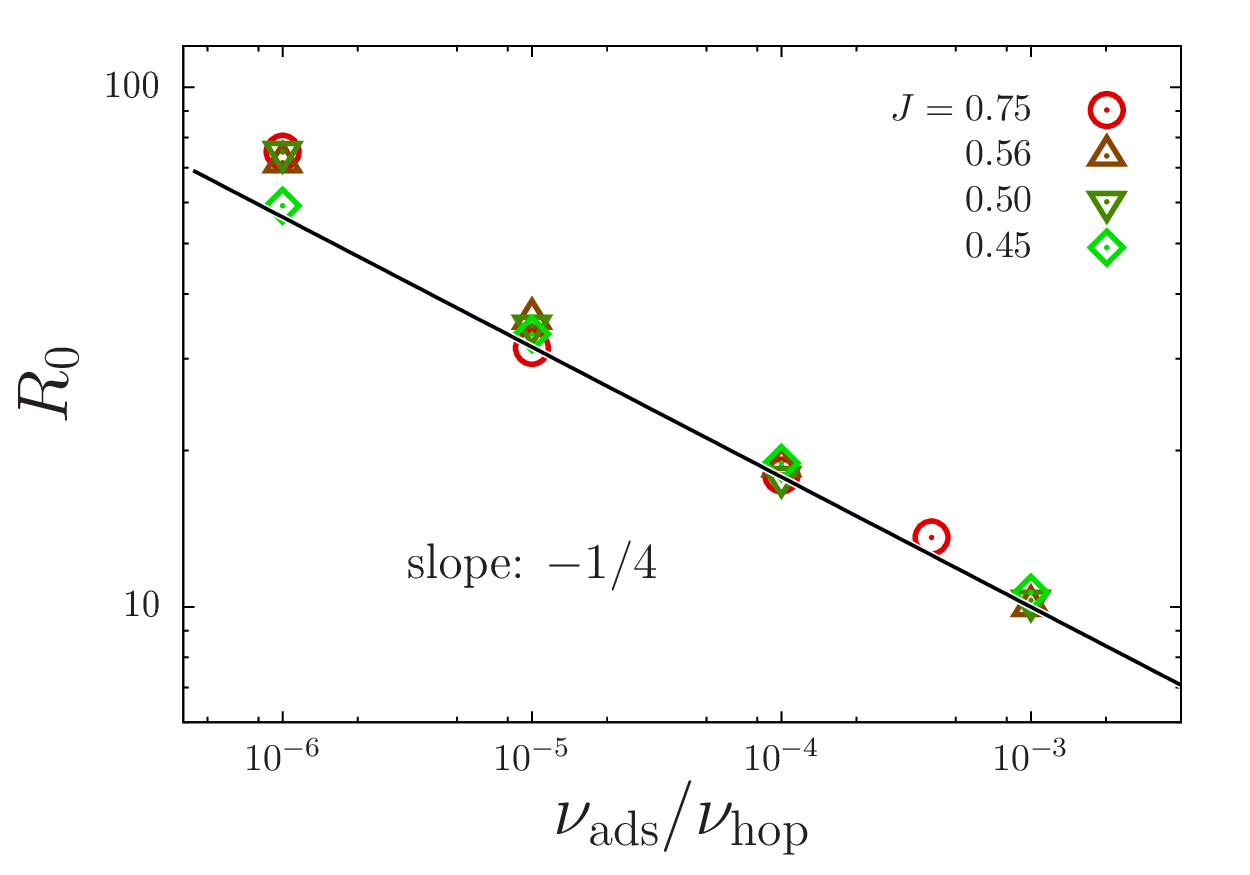}
\caption{Characteristic length scale, $R_0$, of the adlayer structure as a function of the adsorption rate, $\nu_\mathrm{ads}$, for a varying coupling parameter, $J$. The length scale and adsorption rate are given in units of the lattice spacing and hopping attempt frequency, $\nu_\mathrm{hop}$, respectively.}
\label{fig:scaling2}
\end{figure}

The consequence of the deviations from mean-field behavior manifests itself in the feature size of the adlayer structure, $R_0$.
We have defined this length as the parameter to fit $R_\ast = R_0(\nu_\mathrm{ads}t)^{1/3}$ to the simulated data in Figure~\ref{fig:scaling1} at late times.
In Figure~\ref{fig:scaling2} we have plotted $R_0$ as a function of the adsorption rate for various values of the coupling parameter.
The difference with the recently published results is striking: 
Instead of an emerging length scale that decreases with the weak one-sixth power of the adsorption rate \cite{Huston66}, we find a much stronger dependence with a power of approximately one fourth.

This finding implies that local mean-field theories are inadequate in describing the early-stage phase separation of gradually destabilised fluids.
In order to understand how thermal fluctuations affect the structure formation, we set up a simple scaling argument \cite{Buil00, Rullman04}.
We will first apply this approach to retrieve the well-known one-sixth power if phase separation takes place by spinodal decomposition, and subsequently modify this approach to deal with thermal fluctuations.  
For both cases, we start from the realisation that phase separation is entirely governed by diffusion and that the structure must emerge at a time scale 
\begin{equation}
  \tau = L^2(\tau)/D, \label{eq:quench}
\end{equation}
with $L$ the diffusion length and $D$ the diffusivity, which both in principle depend on time due to an ongoing increase of the surface coverage.
Under the presumption that the structures form at a sufficiently low surface coverage, the diffusivity is virtually constant, and is then given by $D=a^2\nu_\mathrm{hop}$ with $\nu_\mathrm{hop}$ the hopping rate.
The task is to estimate how the diffusion length $L$ depends on time, on the hopping rate $\nu_\mathrm{hop}$, and on the adsorption rate $\nu_\mathrm{ads}$.

In case of classical spinodal decomposition following an instantaneous deep quench, the diffusion length is independent of time.
Consequently, short-wavelength density fluctuations decay due to a free energy penalty of density gradients, which is controlled by the gradient stiffness $\kappa$.
The large-wavelength fluctuations are only weakly penalised and are amplified with a rate that, due to the finite rate of mass transport, decreases with an increasing wavelength.
The wavelength that corresponds to the fastest growth is the spinodal length scale, $\lambda_\ast$, and the diffusion length is given by $L\propto \lambda_\ast^2/\sqrt{\kappa}$ \cite{Cahn61}.

In the case that the quench takes place gradually, the spinodal wavelength is infinitely large at the point in time that the spinodal concentration is crossed.
As time proceeds, the fluid is ongoingly destabilised and the spinodal wavelength decreases with the square root of time as $\lambda_\ast(t) \propto ( \nu_\mathrm{hop}t/\kappa)^{-1/2}$.
Inserting this into the expression for the diffusion length and evaluating Eq.~(\ref{eq:quench}) at the quench time $t=\tau$, gives $\tau \propto \nu_\mathrm{hop}^{-1/3}\nu_\mathrm{ads}^{-2/3}$ and $L(\tau)\propto \sqrt{\kappa}(\nu_\mathrm{ads}/\nu_\mathrm{hop})^{-1/6}$ \cite{Huston66, Schaefer15}.

As we have found from our simulations, this result is incorrect due to the influence of thermal fluctuations \cite{Cook70, Langer71}.
Instead of spinodal decomposition through collective diffusion we now deal with nucleation of domains \cite{Lutsko12, Buil00}, which we presume to be facilitated by the self diffusion of individual adsorbates to a cluster.
In this case, the diffusion length is given by the mean-free path length 
\begin{equation}
  L \propto \sqrt{a^2/\theta(t)}, \label{eq:meanfreepath}
\end{equation}
with $a$ again the lattice spacing and $\theta$ the surface coverage. 
Since the surface coverage increases as $\theta(t) \approx \nu_\mathrm{ads}t$, we have $L \propto \sqrt{a^2/\nu_\mathrm{ads}t}$.
After inserting this into Eq.~(\ref{eq:quench}) and evaluating at the quench time $t=\tau$ we obtain the relations $\tau \propto \nu_\mathrm{hop}^{-1/2}\nu_\mathrm{ads}^{-1/2}$ and $L(\tau)\propto a(\nu_\mathrm{ads}/\nu_\mathrm{hop})^{-1/4}$.

This result is consistent with our finding in Figure~\ref{fig:scaling2} and indicates the importance of thermal fluctuations and {diffusion-limited} nucleation in gradually destabilised fluids.
We emphasise that the one-fourth power law is presumably not universal for all gradually quenched systems.
{The first reason for this is that, in practical systems, nucleation may be activation limited, in which case the nucleation rate may be strongly time dependent \cite{Buil00, Rullman04}.}
The second reason is that the structural length scale is determined at the point in time where the rate of fluid structuring outgrows the rate at which disorder is imposed.
In our specific example system, diffusion orders the fluid by cluster formation, and adsorption introduces randomness in the system.
However, the opposite is also possible: If lateral interactions cause adsorption to occur predominantly at the interface of clusters \cite{Kreuzer96, Lausche13}, then adsorption leads to ordering through cluster growth while diffusion could shrink the clusters.
The third reason of why the one-fourth power is presumably not universal follows directly from the ingredients in the same scaling argument: The mean-free path length in Eq.~(\ref{eq:meanfreepath}) depends on the dimensionality of the lattice, i.e., $L \propto a \theta^{-1/d}$.

In summary, we have employed kMC simulations to investigate how the phase separation of molecules adsorbed at a metal surface is affected by ongoing unimolecular adsorption.
Our model system complements previous examples that expose the failure of phenomenological models in surface catalysis \cite{Temel07}.
In our case, this failure originates entirely from nucleation of high-density domains in our adlayer as the unstable region of the phase diagram is gradually entered upon adsorption.
Consequently, the structural length scale is no longer determined by spinodal decomposition through collective diffusion as described by phenomenological models, but by events at the scale of individual molecules.

{These findings have important implications on the interpretation of morphology formation during solvent casting of polymer composites.
Our finding indicates that the mean-field approximation breaks down under conditions where evaporation-induced stratification is absent \cite{Jukes05, Buxton07, Schaefer17}, which arguably takes place under conditions of slow evaporation if a high-boiling-point cosolvent is added to the solution \cite{Cao15, Franeker15B, Franeker15C}.
This failure of the mean-field theory potentially explains the experimental observation of liquid-solid phase-separated morphologies under those conditions \cite{Cao15, Franeker15B, Franeker15C}.}

\begin{acknowledgments}
  The author thanks MSc S. P. Finner, Prof. P. van der Schoot and Dr. A. P. J. Jansen for their critical reading of the manuscript, and acknowledges the Engineering and Physical Sciences Research Council [grant number EP/N031431/1] for partial funding of this project.
\end{acknowledgments}


\end{document}